\documentstyle[aps]{revtex}
\begin{document}
\newcommand{\beq}{\begin{equation}}
\newcommand{\eeq}{\end{equation}}
\newcommand{\bea}{\begin{eqnarray}}
\newcommand{\eea}{\end{eqnarray}}
\def\plumin{\underline{+}}
\def\minplu{{{\stackrel{\underline{\ \ }}{+}}}}
\def\bfr{{\bf r}}
\def\bfw{{\bf w}}
\def\bfc{{\bf c}}
\def\hpsi{\hat \psi (\bfr)}
\def\hpsid{\hat \psi^\dagger (\bfr)}
\def\tpsi{\tilde \psi (\bfr)}
\def\tpsid{\tilde \psi^\dagger (\bfr)}

\title{Excitations of a Bose-condensed gas in anisotropic traps}
\author{D. A. W. Hutchinson and E. Zaremba}
\address{
Department of Physics,
Queen's University,
Kingston, Ontario, Canada K7L 3N6
}

\date{\today}

\maketitle

\begin{abstract}
We investigate the zero-temperature collective excitations of a
Bose-condensed atomic gas in anisotropic parabolic traps. The
condensate density is determined by solving the Gross-Pitaevskii
(GP) equation using a spherical harmonic expansion. The GP
eigenfunctions are then used to solve the Bogoliubov equations
to obtain the collective excitation frequencies and mode densities.
The frequencies of the various modes,
classified by their parity and
the axial  angular momentum quantum number, {\it m},
are mapped out as a function of the axial anisotropy.
Specific emphasis is placed upon the evolution of these
modes from the modes in the limit of an isotropic trap. 

\end{abstract}


\section{Introduction}\label{sect1}
The observation\cite{jin96,mewes96,jin97}
of the collective modes of the Bose condensate in
ultracold trapped atomic gases has stimulated a number of
calculations of the collective excitations in these systems.
Most of these calculations have been performed using the standard 
Bogoliubov equations for {\it T} = 
0\cite{stringari96,edwards96,singh96,java96,castin96,perez96,you97,kagan97}, 
assuming all the atoms 
to be in the condensate, although some finite-temperature 
calculations\cite{hutchinson97,dodd97},
which make use of the Hartree-Fock-Bogoliubov 
equations within the Popov approximation, have also appeared.
Both approaches have been 
used to investigate the excitations in model isotropic traps,
and in anisotropic traps typically
corresponding to the experimental trap of the JILA group
\cite{jin96}.

At the level of the Bogoliubov approximation, the collective
excitations are determined equivalently by solving the coupled
Bogoliubov equations, the linearized time-dependent
Gross-Pitaevskii (GP) equation or a pair of hydrodynamic-like
equations for the condensate density and velocity field.
The methods of solution have included analytical solutions
within the Thomas-Fermi approximation for the 
condensate\cite{stringari96,ohberg97,fliesser97},
variational solutions of the time-dependent GP
equation\cite{castin96,perez96}
and expansion techniques for the solution of the 
Bogoliubov equations using harmonic oscillator 
bases\cite{edwards96,you97}.
In this paper we develop an alternative method of solution
for arbitrary anisotropic traps which is based 
on the construction of the GP equation eigenstates in terms of a
spherical harmonic expansion. The expansion of the Bogoliubov
quasiparticle amplitudes in terms of these functions then leads
to a simplified eigenvalue problem 
which is used to map the collective excitation spectrum
throughout much of the anisotropy parameter space. In
addition, by analyzing the mode densities, we are able to
provide a more detailed discussion than
previously available of the evolution of the modes from the
isotropic limit.

\section{Theory}\label{sect2}
The ground state properties of a trapped atomic Bose gas are
well-represented by the stationary GP equation
\beq
\left [-{\hbar^2 \nabla^2\over 2m} + V_{ext}(\bfr) + gn_c(\bfr)
\right ]
\Phi_0(\bfr) = \mu \Phi_0(\bfr) \,,
\label{GP}
\eeq
where $n_c(\bfr) \equiv |\Phi_0(\bfr)|^2$ is the condensate density
normalized to the total number of particles $N$.
$V_{ext}(\bfr)$ is the 
external confining potential and $g = 4\pi \hbar^2a/m$ is the 
interaction strength determined by the $s$-wave scattering 
length $a$. The ground state eigenvalue $\mu$ is identified with
the chemical potential of the condensate. In the following, we
assume that $V_{ext}(\bfr)$ is axially symmetric,
$V_{ext}(\bfr) = V_{ext}(r,\theta)$, in which case $\Phi_0(\bfr) =
\Phi_0(r,\theta)$. 

Our approach to the solution of the Bogoliubov equations is 
a basis-set expansion method which makes use of the eigenfunctions
of the ground state GP Hamiltonian.
These solutions can be chosen to be 
eigenfunctions of $L_z$ with an angular dependence of $e^{im\phi}$,
where $m$ is an integer. To construct these solutions, we
convert (\ref{GP}) to a matrix problem by making
use of a set of normalized basis functions 
$\psi_{nlm}(\bfr)=R_{nl}(r)Y_{lm}(\hat \bfr)$ which are the
eigenfunctions of the Hamiltonian $\hat h_0 = 
-{\textstyle{\hbar^2 \nabla^2}\over \textstyle{2m}} +
V_0(r)$, where $V_0(r)$ is some spherically symmetric potential. 
One possible choice for this potential
is the spherical average of the external
potential which we assume to be of the harmonic form
\beq
V_{ext}(r,\theta) = {1\over 2}m\omega_{r}^2(x^2+y^2)+
{1\over 2}m\omega_{a}^2z^2\,.
\label{vext}
\eeq
Here, $\omega_r$ and $\omega_a$ are the radial and axial
harmonic frequencies, respectively.
We now expand the external potential in terms of Legendre polynomials
\beq
V_{ext}(r,\theta) = \sum_l V_{ext}^{(l)}(r)P_l(\cos \theta)\,,
\label{exp1}
\eeq
where
\beq
V_{ext}^{(l)}(r) = {2l+1\over 2}\int_0^\pi
d\theta \, \sin \theta P_l(\cos \theta)
V_{ext}(r,\theta).
\label{exp2}
\eeq
The $l = 0$ component is the spherical average $V_{ext}^{(0)}(r)
={1\over 2}mr^2\bar \omega^2$, where $\bar \omega^2 = {1\over 3} 
(2 \omega_r^2 +\omega_a^2)$ is
the arithmetic mean of the squares of the axial and radial frequencies.
The only other term in the expansion of the external potential
is the $l = 2$ component $V_{ext}^{(2)}(r)={1\over 2}mr^2 \beta
\bar \omega^2$ where
\beq
\beta =
{2\omega_{a}^2-2\omega_{r}^2 \over 2\omega_{r}^2+\omega_{a}^2}
\label{beta}
\eeq
defines the anisotropy parameter. It varies
over the range $-1\leq \beta \leq 2$,
where $\beta=-1$ corresponds to the infinitely long cigar-shaped
trap and $\beta=2$ the infinitesimally thin pancake-shaped trap.
The JILA trap\cite{jin96} 
has $\beta = 1.40$ while the MIT trap\cite{mewes96} has $\beta =
-0.991$.

Alternatively, we could use the $l = 0$ component of the total
effective potential $V(\bfr) \equiv V_{ext}(\bfr) + gn_c(\bfr)$
appearing in (\ref{GP}). Regardless of the choice, we define
$V(\bfr) = V_0(r) + \Delta V(\bfr)$ where 
$\Delta V(\bfr)$ is the nonspherical perturbation. (Depending on
the choice of $V_0(r)$, $\Delta V(\bfr)$ may in fact include an
$l = 0$ component.)
Expanding an arbitrary solution of (\ref{GP}) as 
$\phi(\bfr)=\sum_{nlm} a_{nlm} \psi_{nlm}(\bfr)$, the expansion
coefficients are determined by the matrix equation
\beq
(\varepsilon_{nl}-\varepsilon)a_{nlm} + \sum_{n'l'm'} 
\langle nlm|\Delta V|n'l'm'
\rangle a_{n'l'm'} = 0,
\label{mat1}
\eeq
where $\varepsilon$ is a possible eigenvalue and
$\varepsilon_{nl}$ are the eigenvalues of $\hat h_0$. For an axially
symmetric potential,
\beq
\langle nlm|\Delta V|n'l'm'\rangle = \langle nlm|\Delta
V|n'l'm\rangle \delta_{mm'}
\eeq
so that states with different $m$-values remain uncoupled.
However the nonspherical perturbation does have the effect of
coupling basis states with different $l$-values. An
explicit expression for the potential matrix element is
\beq
\langle nlm|\Delta V|n'l'm\rangle = \sum_{\bar l} A_{l\bar l
l'}^m \int_0^\infty dr \, r^2 R_{nl}(r) R_{n'l'}(r) \Delta
V_{\bar l}(r)
\label{element}
\eeq
where $\Delta V_{\bar l}(r)$ are the angular components of
$\Delta V(\bfr)$ defined in analogy with those of $V_{ext}(\bfr)$.
The numerical coefficient in (\ref{element}) is
\beq
A_{l\bar l l'}^m = \sqrt{{2 l'+1 \over 2l+1}} \langle \bar l
l' 0 0 | l 0 \rangle \langle \bar l l' 0 m| lm \rangle
\eeq
where $\langle l_1 l_2 m_1 m_2 | l_3 m_3 \rangle$ is the
usual Clebsch-Gordon coefficient\cite{Messiah}. 
If $\Delta V$ has reflection
symmetry in the $x$--$y$ plane as assumed, only states of the
same parity are coupled ($l$ and $l'$ both 
even or both odd). This restriction implies that the different
$m$-states can be chosen to have a well-defined parity 
$\Pi = (-1)^{l+m}$ with respect to reflections in the $x$--$y$ plane.

The condensate wave function, $\Phi_0(\bfr) = \sqrt{N}
\phi_0(\bfr)$, is determined by the lowest energy
even-parity solution of (\ref{mat1}) 
in the $m=0$ subspace and is given by
\beq
\Phi_0(r,\theta)=\sqrt{N} \sum_{nl}a_{nl0}^{(0)}R_{nl}(r)Y_{l0}(\bfr)
\,.
\label{gs}
\eeq
We can now evaluate the condensate density $n_c(r,\theta)$, which
has an expansion similar to that of the potential:
\beq
n_c(r,\theta)=\sum_l n_l(r)P_l(\cos \theta)\,.
\label{nc}
\eeq
Comparing (\ref{nc}) to the square of (\ref{gs}), we obtain
\beq
n_{\bar l}(r) = {N \over 4\pi}
\sum_{{nl \atop n'l'}} \sqrt{(2l+1)(2l'+1)}
|\langle l l' 00 | \bar l 0 \rangle|^2 a_{nl0}^{(0)}
a_{n'l'0}^{(0)} R_{nl}(r) R_{n'l'}(r)\,.
\eeq
These radial functions provide what is needed to complete the
specification of the nonspherical potential $\Delta V$. 
Since Eq.(\ref{mat1}) depends on the condensate density through
$\Delta V$, this equation
must be iterated until a self-consistent solution
for the condensate wave function is generated.

We determine the collective excitations of the condensate using
the Bogoliubov equations, which are equivalent to solving the
linearized time-dependent GP equation. As shown in 
Ref.\cite{hutchinson97}, the
Bogoliubov equations can be cast into the form
\bea
\hat h^2 \psi_i^{(-)}(\bfr) + 2g\hat h n_c(\bfr)
\psi_i^{(-)}(\bfr) &=& E_i^2 \psi_i^{(-)}(\bfr)
\nonumber \\
\hat h^2 \psi_i^{(+)}(\bfr) +2gn_c(\bfr) \hat h
\psi_i^{(+)}(\bfr) &=& E_i^2 \psi_i^{(+)}(\bfr)\,, \label{psi+}
\label{9}
\eea
where the functions $\psi_i^{(\pm)}$ are related to the
quasiparticle amplitudes by $\psi_i^{(\pm)}(\bfr) \equiv
u_i(\bfr) \pm v_i(\bfr)$. The Hamiltonian $\hat h$ appearing
in these equations is the ground state condensate Hamiltonian
shifted by the ground state eigenvalue $\mu$. Since these equations
are uncoupled, either can be used to determine the excitation
energies $E_i$.

To solve (\ref{9}) for $\psi_i^{(+)}(\bfr)$ we proceed as in
Ref.\cite{hutchinson97}
and introduce the expansion $\psi_i^{(+)}(\bfr) =
\sum_\alpha c_\alpha^{(i)} \phi_\alpha(\bfr)$ where the
$\phi_\alpha(\bfr)$ are the eigenfunctions of $\hat h$
determined from (\ref{mat1}), and the expansion coefficients
$c_\alpha^{(i)}$ are required to be normalized as $\sum_\alpha
\varepsilon_\alpha c_\alpha^{(i)*} c_\alpha^{(j)} =
E_i\delta_{ij}$. Substituting this expansion into the equation
for $\psi_i^{(+)}(\bfr)$, we obtain
\beq
\sum_\beta \left \{ M_{\alpha \beta} + \varepsilon_\alpha
\delta_{\alpha \beta} \right \} \varepsilon_\beta c_\beta^{(i)}
= E_i^2 c_\alpha^{(i)}\,.
\label{12}
\eeq
Within a particular $m$-subspace, the
matrix $M_{\alpha \beta}$ is given by 
\bea
M_{\alpha \beta}&=&2g \int d\bfr \,\phi^*_\alpha(\bfr)
n_c(\bfr)\phi_\beta(\bfr)
\nonumber \\
&=&\sum_{nln'l'}a_{nl}^{(\alpha)*}a_{n'l'}^{(\beta)}
\langle nlm|2gn_c| n'l'm \rangle
\label{mmat}
\eea
where $a_{nl}^{(\alpha)}$ is the $\alpha$-th eigenvector of
Eq.(\ref{mat1}). Since the matrix $M_{\alpha \beta}$ is
diagonal in both the azimuthal quantum number $m$ and the parity
$\Pi$, these quantum numbers also serve to classify the
collective modes. Once the eigenvector $c_\alpha^{(i)}$ has been
determined, the density of the $i$-th mode is given by
\beq
\delta n_i(\bfr) \propto \phi_0(\bfr) \psi_i^{(-)}(\bfr) =
\phi_0(\bfr) \sum_\alpha {\varepsilon_\alpha \over E_i}
c_\alpha^{(i)} \phi_\alpha(\bfr)\,.
\label{density}
\eeq

\section{Results}\label{sect3}
As an illustration of the technique we consider the case of 2000
rubidium atoms contained  within an axially symmetric
harmonic trap of varying
anisotropy. For the JILA trap, $\omega_r/2\pi = 75$ Hz and
$\omega_a/2\pi = 212$ Hz, giving an anisotropy parameter of
$\beta=1.40$, and an averaged harmonic frequency of $\bar
\omega/2\pi =137$ Hz. We keep the latter fixed in our
calculations and vary the parameter $\beta$. To complete the
parameter specification, we take $m(^{87}$Rb$) = 1.44 \times
10^{-25}$ kg for the mass of the atoms, and an $s$-wave
scattering length of $a \simeq 110 a_0 = 5.82 \times 10^{-9}$ m. 
Throughout, lengths and energies are expressed in terms of 
the characteristic oscillator
length of the isotropic trap $d = (\hbar/m \bar \omega)^{1/2} = 9.21
 \times 10^{-7}$ m and the characteristic trap energy $\hbar \bar \omega
= 9.03 \times 10^{-32}$ J, respectively. When expressed in these
units, the dimensionless condensate wave function depends only
on the dimensionless parameters $\beta$ and $\gamma \equiv Na/d$.

In Fig. 1 we show  the excitation spectrum obtained from the
solution of
(\ref{12}). In these calculations the basis functions, generated
numerically, are limited in number to
between 100 and 200, the actual
number being controlled by the value of the high-energy cutoff
set in the basis function expansion. This number provided
sufficient accuracy over the range of the anisotropy parameter
shown. However as $\beta$ increases towards its extreme values
of $-1$ and 2, increasingly more basis functions are required to
obtain accurate results and our calculations are necessarily
curtailed.

The modes shown here are those corresponding to axial quantum numbers
$m=0-4$ for both even and odd parity. The modes of even parity are
shown in Fig. 1(a) and those of odd parity in Fig. 1(b). 
Let us first examine the
modes of the isotropic trap along the line $\beta=0$ studied
previously\cite{singh96,hutchinson97}. 
In order of increasing frequency, we find a doubly
degenerate mode corresponding to the $l=1$ excitations, followed
by the triply degenerate $l=2$ mode, the quadruply degenerate
$l=3$ mode and a nondegenerate $l=0$ mode. (In this description,
we do {\it not} include the additional degeneracy associated with the
sign of $m$ which is not lifted by axial anisotropy.) 
It is on the evolution
of these modes that we concentrate in the following.

The lowest doubly degenerate $l=1$ modes for the spherical trap 
correspond to the centre of mass modes
of the harmonic potential\cite{dobson94}. 
At $\beta=0$ both modes have a frequency of exactly 1 in
units of $\bar \omega$. Of these two modes one is the odd parity, 
$m=0$ mode, while the other is the even parity $m =1$ mode. (Recall
that parity refers to the reflection symmetry in the $x$--$y$ plane.)
Both of
these modes correspond to a rigid oscillation of the condensate
density in the axial and transverse directions, respectively. 
As $\beta$ begins to deviate from zero, the degeneracy of these
two modes is lifted, one oscillating at the frequency 
$\omega_a$ (odd parity) and the other at $\omega_r$ (even
parity). When expressed in units of $\bar \omega$, 
these modes disperse according to
$\omega_a/{\bar \omega} = \sqrt{1+\beta}$ and $\omega_r/{\bar
\omega} = \sqrt{1-\beta/2}$. We have plotted these exact results
for the center of mass modes in Fig. 1 to illustrate the accuracy 
of our numerical calculations. 
Any deviation of the numerical results from the exact values
would indicate the need to increase
the number of basis functions in the expansions.

For the experiments in the regime where the axial confining potential 
is stronger than in the radial direction (such as for the JILA trap),
the collective modes of interest, other than the centre of mass modes,
are those originating from the $l=2$ mode in the isotropic limit. In
particular, the modes observed experimentally\cite{jin96} are  
the $m=0$ and $m=2$ even parity modes originating from the $l=2$ mode, 
the lower frequency mode being the $m=2$.
This latter mode has quadrupolar character, with a density fluctuation 
which is concentrated in the $x$--$y$ plane. As a result, the
oscillation is mainly influenced by the radial trap frequency
$\omega_r$ and the mode frequency decreases monotonically with
increasing $\beta$, similar to the behaviour of the $m = 1$
centre of mass mode. In fact, as will be explained in more
detail shortly, these particular $m=1$ and $m=2$ modes are the first two
members of a family of modes identified by the number `1' in
Fig. 1(a).

We now examine the even parity $m=0$ mode originating from
$l=2$ in more detail. The
mode density defined by (\ref{density})
is shown in Figs. 2 and 3 for a variety of $\beta$
values. Since the mode density is of the form $\delta n({\bf r})
\propto f(r,\theta)$, the interesting spatial dependence is
revealed by considering a plane (e.g. the $x$--$z$ plane)
containing the axis of the trap. Fig. 2 gives the behaviour of
the density along the $x$ and $z$ directions while Fig. 3 gives
a contour representation. In the isotropic limit, this mode
corresponds to the quadrupole $l=2$, $m=0$ mode for which an
expansion (contraction) in the radial direction is accompanied
by a contraction (expansion) in the axial direction, as shown in
the $\beta =0$ panel of Fig. 2. The contour representation in
Fig. 3(b) shows nodal lines making angles of $\pm 54.7^\circ$ with
respect to the $z$-axis. This representation is particularly
useful since it indicates the direction of particle flow. In the
Thomas-Fermi approximation\cite{stringari96}, the velocity field
is given by $m{\textstyle{\partial {\bf v}}
 \over \textstyle{\partial t}} = -g
\nabla\delta n$ which implies that the direction of particle
flow is normal to the density contours. Although not exact,
this relationship between velocity and mode density is still a 
good guide in the Bogoliubov approximation. Fig. 3(b) thus
indicates that for this mode there is a circumferential flow
from the equatorial $x$--$y$ plane to the polar regions.

However, as the magnitude of $\beta$ is increased from zero, the
character of the mode changes from quadrupolar to one which is
more accurately described as `breathing-like'. For negative
$\beta$, the cloud breathes in the axial direction, while for
positive $\beta$, it breathes in the radial direction.
Thus, in Fig. 3(a) for $\beta = -0.5$, we see nodal surfaces
which are approximately planes
perpendicular to the $z$-axis, indicating a flow of atoms in the
axial direction. This mode is analogous to the lowest standing
wave resonance in an open-ended organ pipe of length $L$, with
wavelength $\lambda \simeq L$.
Conversely, for $\beta = 1.4$ in Fig. 3(c), we
see a cylindrical nodal surface and a flow which is
predominantly radial. In view of this behaviour, the mode
frequency would be expected to depend mainly upon $\omega_a$ for
negative $\beta$ and upon $\omega_r$ for positive $\beta$,
resulting in a dispersion to zero in the $\beta \to -1$ and
$\beta \to 2$ limits. This is precisely the behaviour exhibited
by this mode which is the lowest mode labeled `3' in Fig. 1(a).
Similar arguments can be used to understand the dispersion of
the higher frequency modes once a contour representation of the
mode density is available. For example, we show in Figs. 4(a-c)
the contour representation of the lowest odd-parity mode labeled
`4' in Fig. 1(b), which originates from the $l=3$, $m=0$ mode in
the isotropic limit. Fig. 4(a) for $\beta = -0.5$ is clearly the
next standing-wave resonance following the fundamental mode
illustrated in Fig. 3(a). Likewise, Fig. 4(c) for $\beta = 1.4$
illustrates a hybrid mode which involves both the radial
breathing motion of Fig. 3(c) and an oscillation in the axial
direction. As a result of this axial motion, this mode has a
finite limiting frequency for $\beta \to 2$.
 
Also of interest within the anisotropy regime of the JILA experiment 
is the $m=3$ mode with even parity, originating from  the $l=3$ mode 
for the isotropic case (the third curve from the bottom,
labeled `1' in Fig. 1(a)). The excitation energy of this mode is 
only slightly greater than that of  the experimentally
observed $m=0$ mode and like the latter, disperses to zero for
$\beta \to 2$. However, as indicated by the common label of `1'
in Fig. 1(a), this mode is the third member of the family
mentioned previously, which is distinct from the family labeled `3'
that the $m=0$ mode belongs to. The observability of this $m=3$ 
mode would of course depend on the possibility of inducing 
experimentally an
excitation having a $e^{3i\phi}$ azimuthal dependence.

Although the frequency spectrum in Fig. 1 as a function $\beta$
is quite complex, it is apparent that certain patterns have
emerged. These correspond to the families labeled by 1 through 4 
in the figure, three of which were alluded to previously.
These families correspond to the modes very recently calculated 
analytically in the Thomas-Fermi limit by \"Ohberg et
al.\cite{ohberg97} and independently by
Fliesser et al.\cite{fliesser97}. (Some of these mode
frequencies were obtained earlier by
Stringari\cite{stringari96}.)
Their analytic expressions for
the mode frequencies of the four families, converted
to our notation, with $\omega_i$ 
corresponding to the ${\it i}$-th group as labeled in Fig. 1,
are given below: 
\bea
\omega_1^2 &=& (1-\beta/2)m \bar \omega^2
\nonumber \\
\omega_2^2 &=& [(1+\beta)+(1-\beta/2)m]\bar \omega^2
\nonumber \\
\omega_3^2 &=&  \Bigg [{3\over2}(1+\beta)+(m+1)(2-\beta)
\nonumber \\
&-& 1/2  \left (9(1+\beta)^2-2(m+4)(2-\beta)(1+\beta)
+(m+2)^2  (2-\beta)^2  \right )^{1/2} \Bigg ] \bar \omega^2
\nonumber \\
\omega_4^2 &=& \Bigg[{7\over 2}(1+\beta)+(m+1)(2-\beta) 
\nonumber \\
&-& 1/2  \left (25(1+\beta)^2+2(m-4)(2-\beta)(1+\beta)+(m+2)^2 
(2-\beta)^2  \right )^{1/2}\Bigg ] \bar \omega^2\,.
\label{analytic}
\eea
In the notation of Fliesser et al.\cite{fliesser97}, these modes
are labeled by three quantum numbers $(n,j,m)$, where $m$ is
the usual azimuthal quantum number and $n$ and $j$ are two
others which distinguish the radial and angular character of the
modes. The four families we have identified correspond to
(0,0,$m$), (1,0,$m$), (2,0,$m$) and (3,0,$m$), respectively.
This identification can be checked by considering the limiting
values of the analytic result for $\beta \to -1$ and $\beta \to
2$, as presented in Table 1. For example, the 1-modes in Fig.
1(a) tend to zero as $\beta \to 2$ and to a finite value for
$\beta \to -1$, in accord with the analytic results. However the
finite limiting values of the numerical results deviate from the
analytic value of $\sqrt{3m/2}$, the difference increasing with
increasing $m$ (the agreement is exact for the lowest mode in
this family since it corresponds to the center of mass mode). The
discrepancy between the numerical and analytic results is not
due to inaccuracies in the numerical calculations, since we have
checked that the numerical basis set is sufficiently large to
yield accurate results for the modes shown. Rather, the
differences are real and reflect the fact that our calculations
are for a finite number of particles, $N$, while the
Thomas-Fermi results correspond to the $N \to \infty$ limit. 
Similar deviations
appear for the 2-modes in Fig. 1(b). The analytic limits for
these modes are $\sqrt{3}$ for $\beta \to 2$ and $\sqrt{3m/2}$
for $\beta \to -1$. The lowest 2-mode in Fig. 1(b), the 
odd-parity center of mass mode, does extrapolate to $\sqrt{3}$
but the higher modes in this family, to within our numerical
accuracy, do not. In other words, the finite size of the trapped
gas leads to a weak $m$-dependence of the limiting value, 
which presumably disappears as $N \to \infty$.
As noted by Fliesser et al.\cite{fliesser97}, the
convergence with $N$ is more rapid for modes having smaller $n$
values, which is consistent with our numerical findings.

Another feature of the mode spectrum worth noting is 
a number of instances of
anticrossing-type behaviour between modes of equal $m$ and the
same parity. The most obvious example of this occurs in Fig.
1(a) for the two $m=0$ modes near $\beta =1$. Similar behaviour
is seen for two of the $m =0$  odd-parity modes in Fig. 1(b) near  
$\beta = -0.5$. Of course, there is no avoided crossing of modes
with different $m$ values, and there are numerous examples of
these crossings in Fig. 1. However for real traps which do not
possess ideal cylindrical symmetry, one can expect to see
additional anticrossings associated with the coupling between
modes induced by trap imperfections.

\section{Conclusions}\label{sect4}
In conclusion, we have presented a new and efficient  method for  
solving the Bogoliubov equations for a gas of  weakly interacting Bose
condensed atoms in an axially symmetric magnetic trap. We have used the 
method to calculate
the low lying collective mode frequencies and densities over a 
large portion of the anisotropy parameter space. The
evolution of the modes as a function of the anisotropy parameter
$\beta$ has provided a more complete understanding of the
relationship of the modes in anisotropic traps to those in the
isotropic limit. We have also made contact with analytical
results obtained in the Thomas-Fermi limit and have identified 
differences between the finite-$N$ and infinite-$N$
calculations. We are presently applying our method for anisotropic
traps to the problem of finite-temperature excitations in the 
hydrodynamic  regime\cite{zaremba97}.

\acknowledgements
This work was supported by grants from the Natural Sciences and
Engineering Research Council of Canada.

\bigskip
\begin{table}    {\small
Table 1: Limiting values of the mode frequencies, 
$\omega_i/\bar \omega$, given in Eq.(\ref{analytic}).}

\begin{tabular}{ccc}                  
  i     & $\beta \to -1$  &  $\beta \to 2$  \\ 
\hline  
1 & $\sqrt{{3\over 2} m}$ & 0 \\
2 & $\sqrt{{3\over 2} m}$ & $\sqrt{3}$ \\
3 & $\sqrt{{3\over 2} m}$ & 0 \\
4 & $\sqrt{{3\over 2} m}$ & $\sqrt{3}$ 
\end{tabular} 
\end{table}  
\begin{itemize}
\vfil \break
\centerline{FIGURE CAPTIONS}
\item[Fig.1:]
Mode frequencies in units of the average trap frequency $\bar
\omega$ as a function of the anisotropy parameter $\beta$: (a)
even-parity modes and (b) odd-parity modes. The symbols code the
various $m$-values: $m=0$ (open circle), $m=1$ (filled square),
$m=2$ (open triangle), $m=3$ (filled circle), $m=4$ (open
square). The numerical labels 1--4 identify the different
families of modes as discussed in the text.
 
\item[Fig.2:]
Mode density (left panel) and corresponding equilibrium density
(right panel) for values of the anisotropy parameter shown. The
solid line gives the variation along the $x$-axis and the chain
curve along the $z$-axis.
 
\item[Fig.3:]
Contour representation of the mode density of the
lowest mode ($m=0$, even parity)
labeled `3' in Fig. 1(a): (a) $\beta = -0.5$, (b) $\beta = 0$
and (c) $\beta = 1.4$. The shaded region corresponds to negative
values of the mode density, the unshaded region to positive
values.
 
\item[Fig.4:]
As in Fig. 3, but for the lowest mode ($m=0$, odd parity) labeled
`4' in Fig. 1(b).

\end{itemize}


\begin{references}
\bibitem{jin96} D. S. Jin, J.R. Ensher, M.R. Matthews, C.E.
Weiman and E.A. Cornell, Phys. Rev. Lett. {\bf 77}, 420 (1996).
\bibitem{mewes96} M.-O. Mewes, M.R. Andrew, N.J. van Druten,
D.M. Kurn, D.S. Durfee, C.G. Townsend and W. Ketterle,
Phys. Rev. Lett. {\bf 77}, 988 (1996).
\bibitem{jin97} D. S. Jin, M.R. Matthews, J.R. Ensher, C.E.
Weiman and E.A. Cornell, Phys. Rev. Lett. {\bf 78}, 764 (1997).
\bibitem{stringari96} S. Stringari, Phys. Rev. Lett. {\bf 77},
2360 (1996).
\bibitem{edwards96} M. Edwards, P.A. Ruprecht, K. Burnett, R.J.
Dodd and C.W. Clark, Phys. Rev. Lett. {\bf 77}, 1671 (1996).
\bibitem{singh96} K. G. Singh and D. S. Rokhsar, Phys. Rev.
Lett. {\bf 77}, 1667 (1996).
\bibitem{java96} J. Javanainen, Phys. Rev. A {\bf 54}, 3722
(1996).
\bibitem{castin96} Y. Castin and R. Dum, Phys. Rev. Lett. 
{\bf 77}, 5315 (1996).
\bibitem{perez96} V.M. P\'erez-Garc\'\i a, H. Michinel, J.I. Cirac,
M. Lewenstein and P. Zoller, Phys. Rev. Lett. {\bf 77}, 5320 (1996).
\bibitem{you97} L. You, W. Hoston, and M. Lewenstein, Phys. Rev.
A {\bf 55}, 1581 (1997).
\bibitem{kagan97} Yu. Kagan, E.L. Surkov and G.V. Shlyapnikov,
Phys. Rev. A {\bf 55}, R18 (1997).
\bibitem{hutchinson97} D.A.W. Hutchinson, E. Zaremba and A. Griffin,
Phys. Rev. Lett. {\bf 78}, 1842 (1997).
\bibitem{dodd97} R.J. Dodd, M. Edwards, C.W. Clark and K. Burnett,
preprint.
\bibitem{ohberg97} P. \"Ohberg, E.L. Surkov, I. Tittonen, S.
Stenholm, M. Wilkens and G.V. Shlyapnikov, preprint
physics/9705006.
\bibitem{fliesser97} M. Fliesser, A. Csord\'as, P. Sz\'epfalusy
and R. Graham, preprint cond-mat/9706002.
\bibitem{Messiah} A. Messiah, {\it Quantum Mechanics} (North
Holland, Amsterdam, 1966), vol. 2.
\bibitem{dobson94} See J.F. Dobson, Phys. Rev. Lett. {\bf 73},
2244 (1994) and references therein for a discussion of the
center of mass mode in parabolic traps and the generalized Kohn
theorem.
\bibitem{zaremba97} E. Zaremba, A. Griffin and T. Nikuni,
preprint cond-mat/9705134.

\end{references}
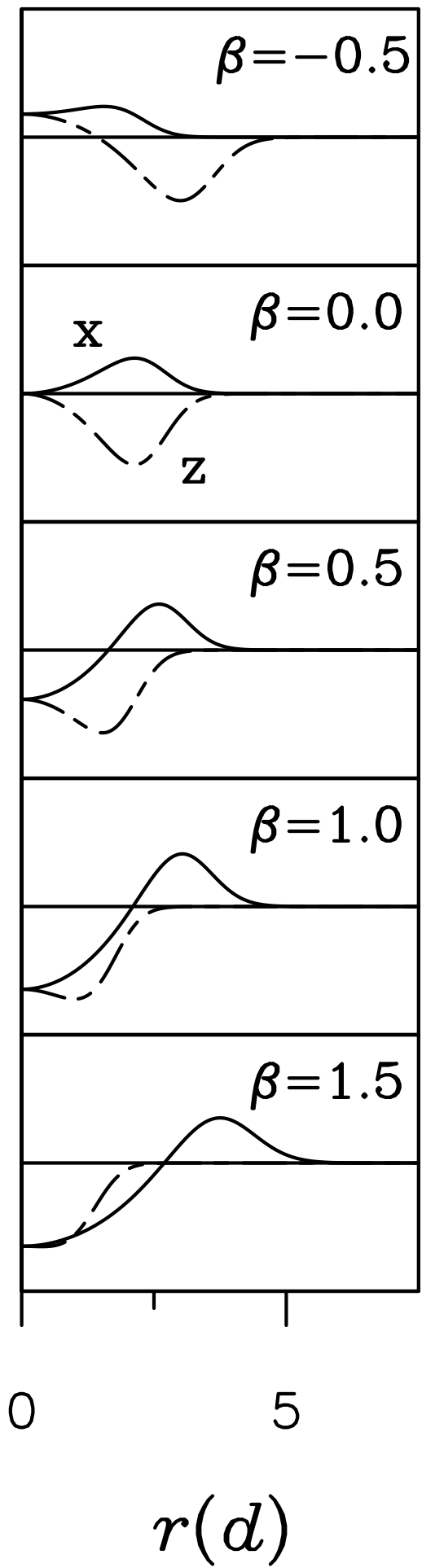
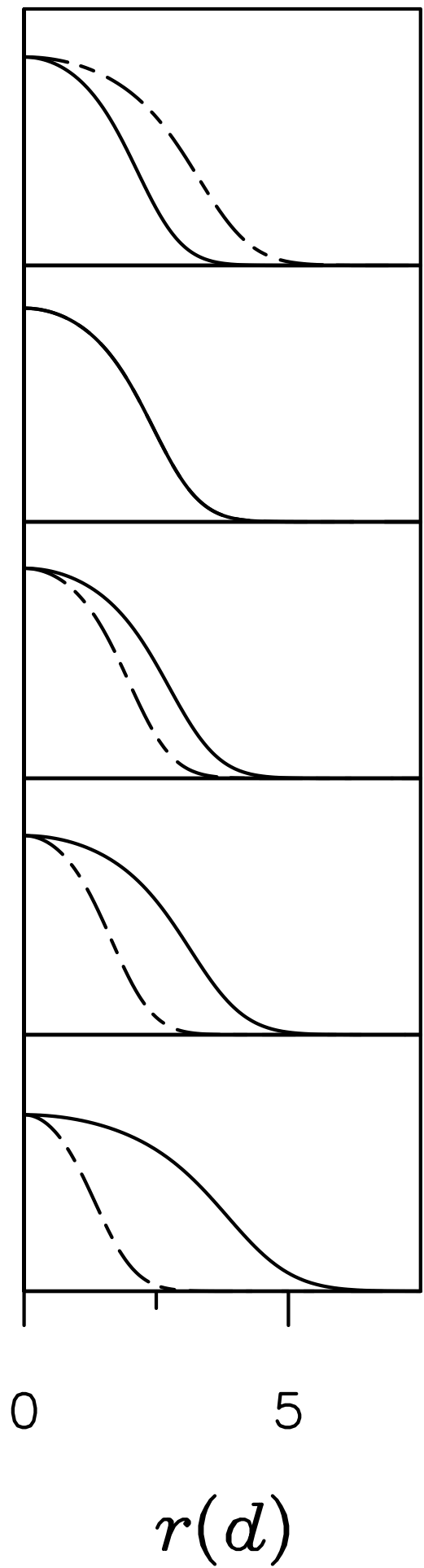
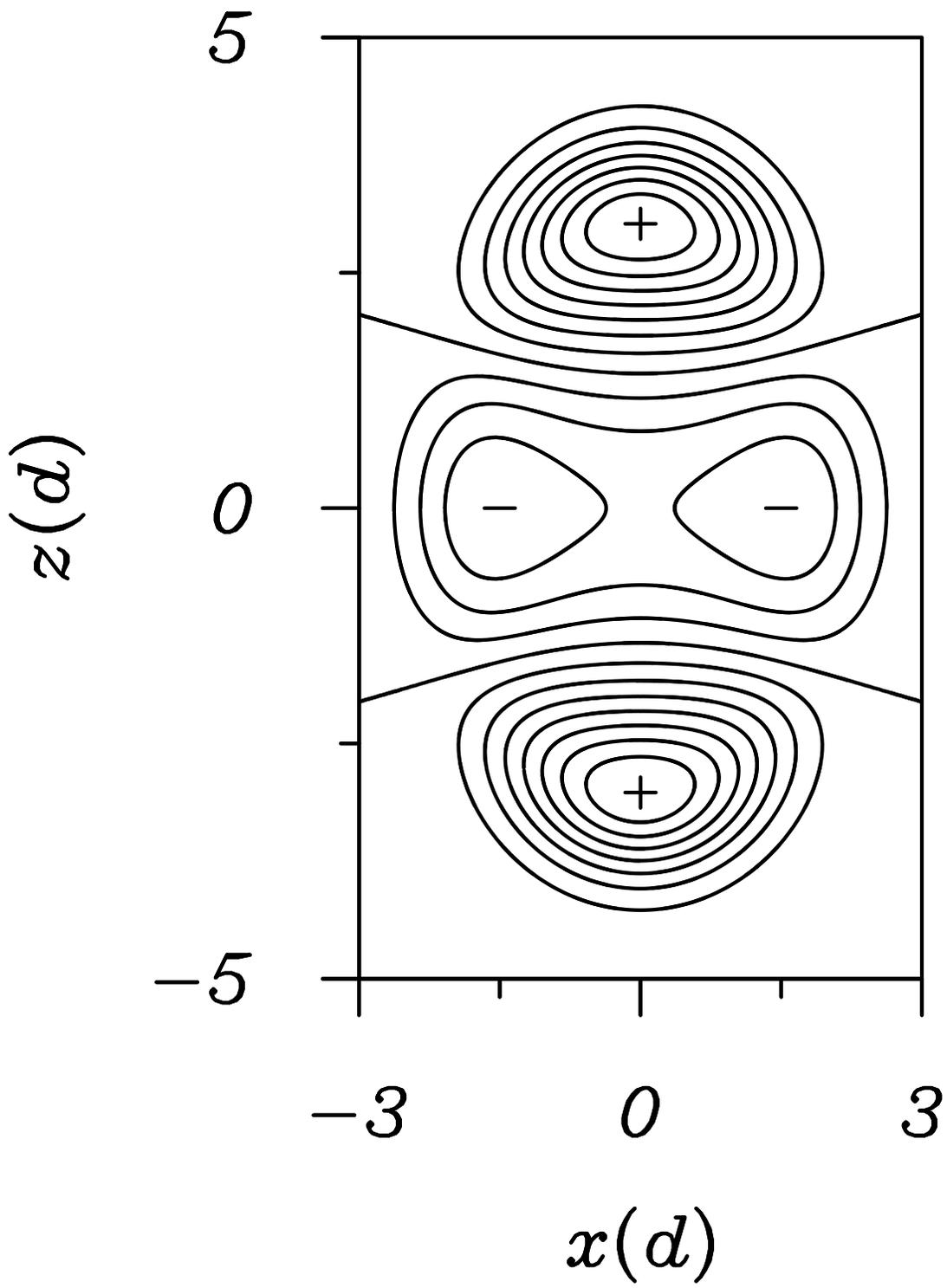
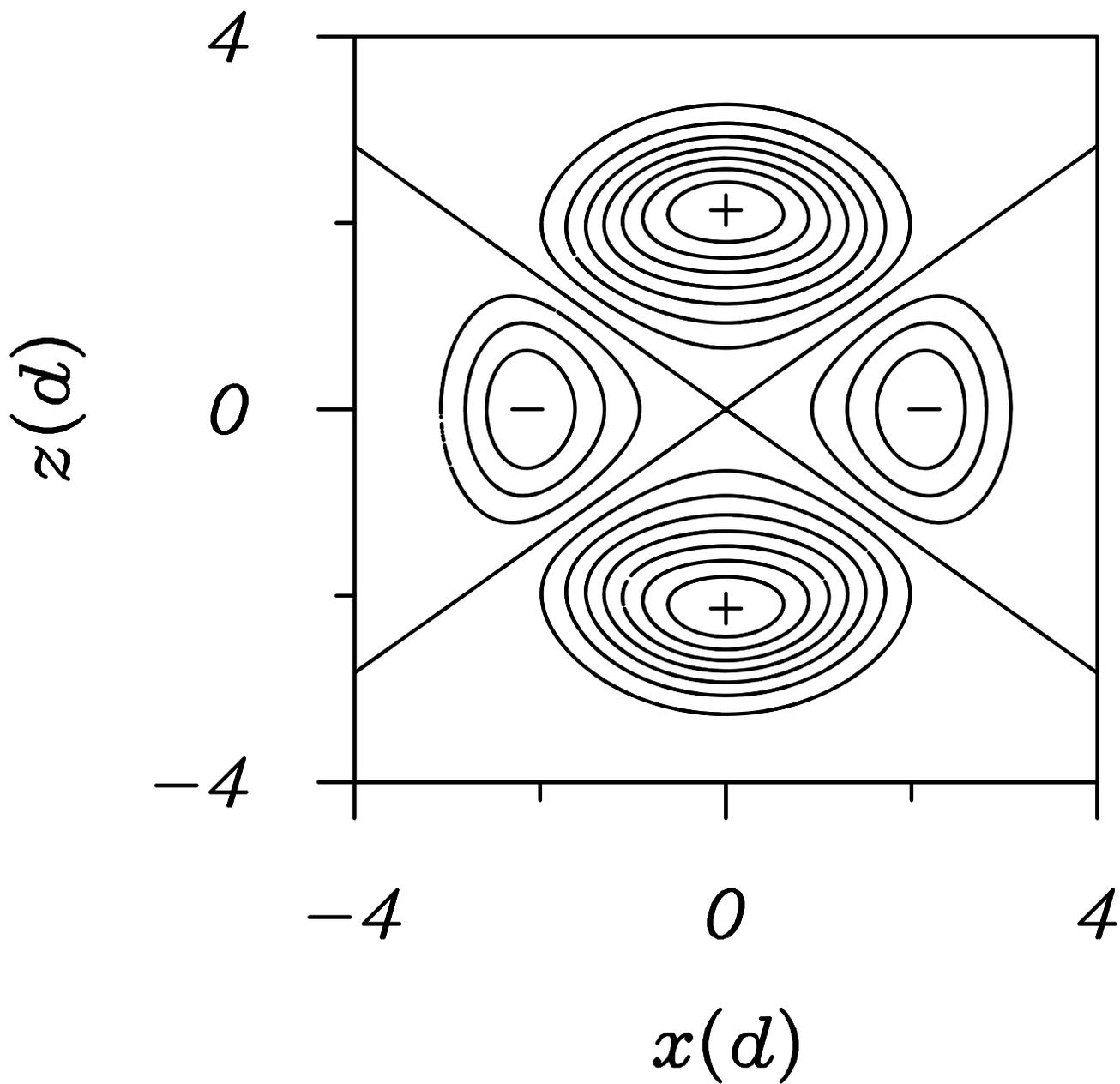
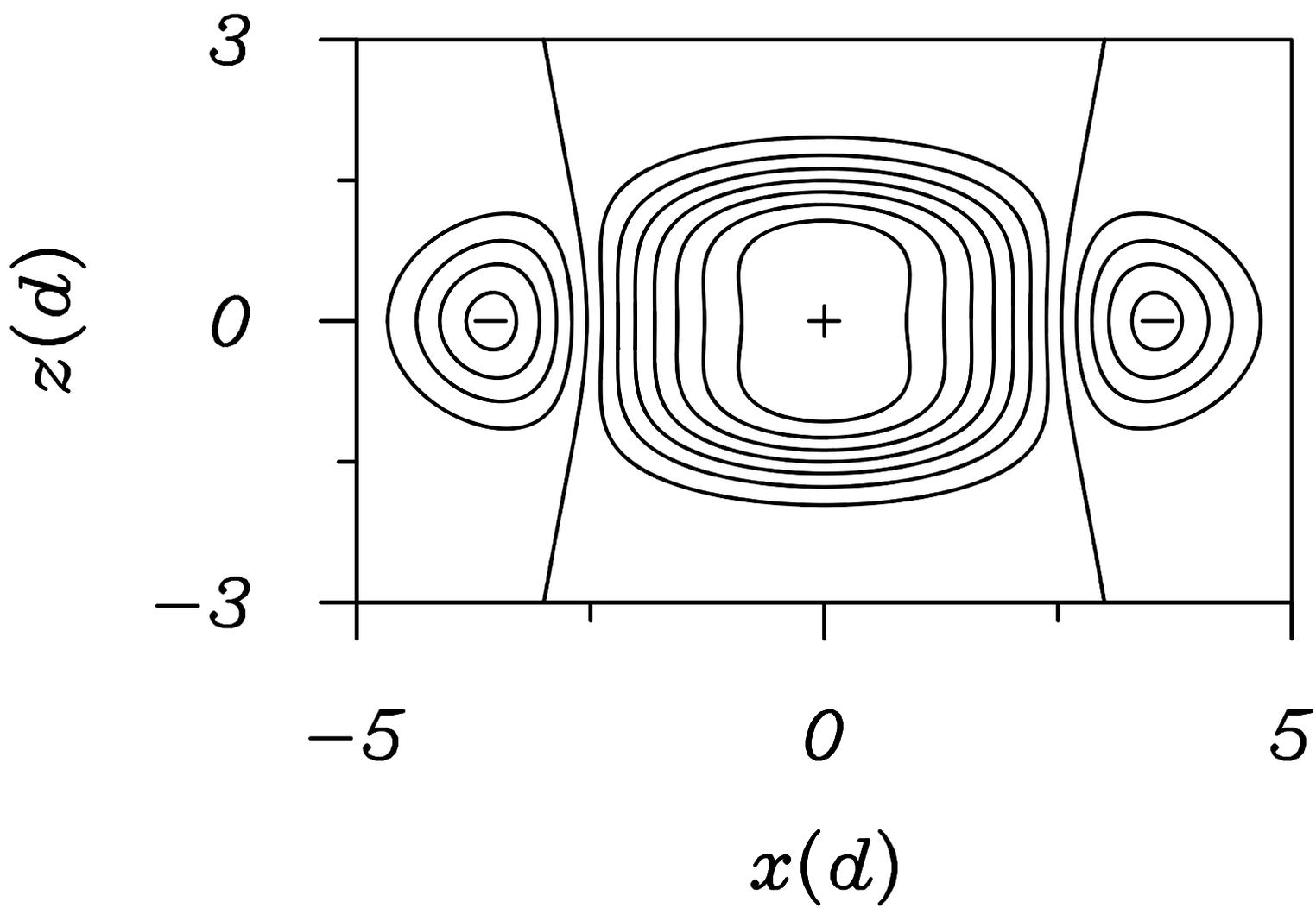
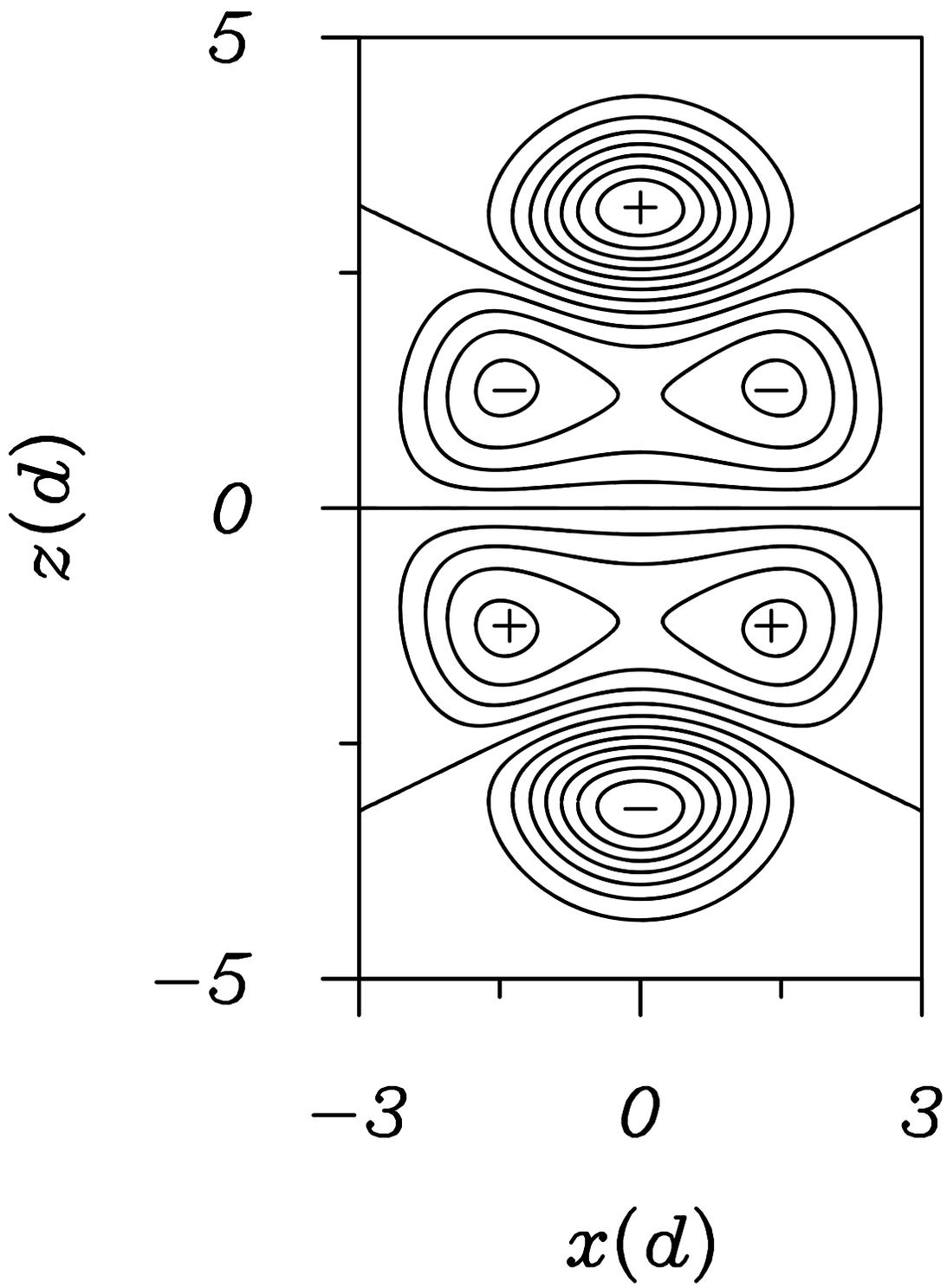
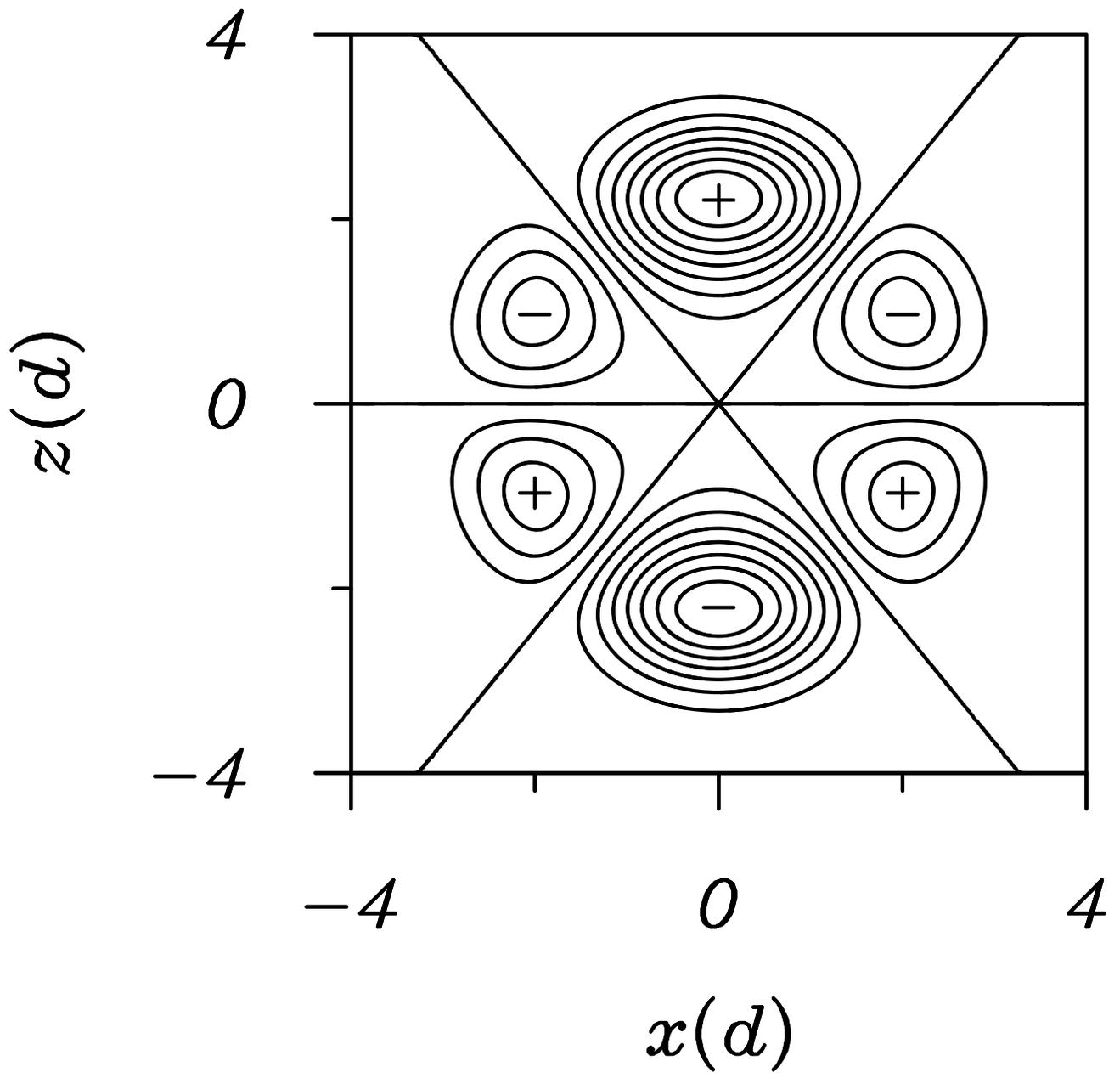
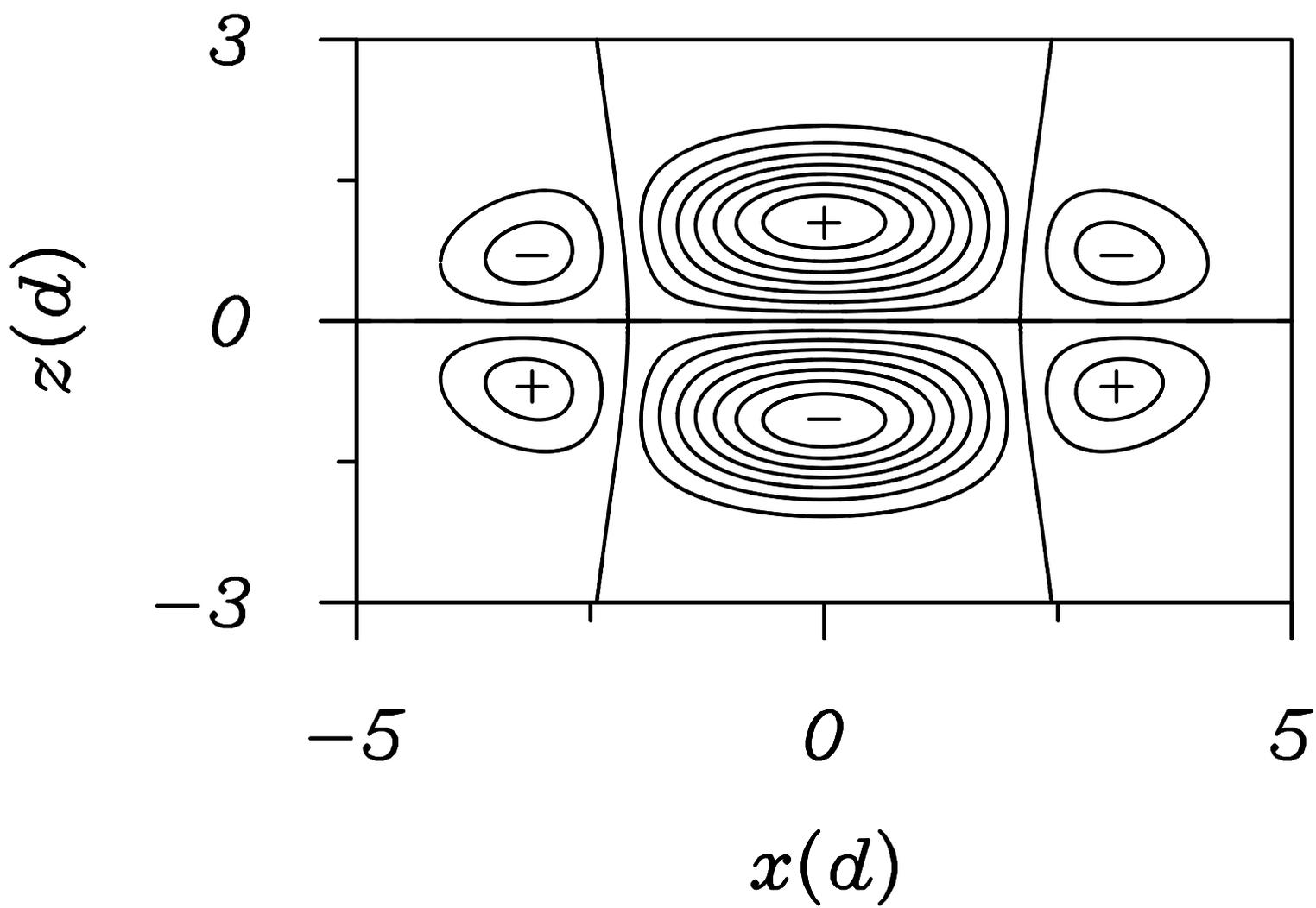
\end{document}